# Uncovering Regulatory Affairs Complexity in Medical Products: A Qualitative Assessment Utilizing Open Coding and Natural Language Processing (NLP)

Yu Han[*1], Aaron Ceross[†1], and Jeroen H.M. Bergmann[‡1]

[1]*Department of Engineering Science, University of Oxford, Old Road Campus, Oxford, OX3 7QD, United Kingdom*

December 30, 2023

**Abstract**

This study investigates the complexity of regulatory affairs in the medical device industry, a critical factor influencing market access and patient care. Through qualitative research, we sought expert insights to understand the factors contributing to this complexity. The study involved semi-structured interviews with 28 professionals from medical device companies, specializing in various aspects of regulatory affairs. These interviews were analyzed using open coding and Natural Language Processing (NLP) techniques. The findings reveal key sources of complexity within the regulatory landscape, divided into five domains: (A) Regulatory language complexity, (B) Intricacies within the regulatory process, (C) Global-level complexities, (D) Database-related considerations, and (E) Product-level issues. The participants highlighted the need for strategies to streamline regulatory compliance, enhance interactions between regulatory bodies and industry players, and develop adaptable frameworks for rapid technological advancements. Emphasizing interdisciplinary collaboration and increased transparency, the study concludes that these elements are vital for establishing coherent and effective regulatory procedures in the medical device sector.

# Keywords

[*]yu.han@eng.ox.ac.uk (Corresponding Author)
[†]aaron.ceross@eng.ox.ac.uk
[‡]jeroen.bergmann@eng.ox.ac.uk

# 1  Introduction

Regulatory affairs, crucial in industries such as healthcare, involves understanding and adhering to regulations to ensure product safety, efficacy, and quality requirements are met while promoting innovation [1]. Understanding the complexity of this work is essential due to legal implications, its impact on public health, and the evolving regulatory landscape. Moreover, regulatory affairs play a vital role in safeguarding public health, monitoring post-market performance, and adapting to evolving guidelines. Constant awareness of regulatory changes is crucial for professionals and policymakers, fostering evidence-based policy-making. Investigating regulatory affairs is indispensable for legal compliance, public health impact, and adapting to evolving industry needs [2].

In the process of bringing a medical device product to the market, it is necessary to obtain approval from the Medical Device Regulatory Authority (MDRA). The MDRA rigorously reviews the device dossier to determine its compliance with safety and efficacy requirements outlined in regulations and guidance. Regulations, as noted by Sakuma [3], can act as a double-edged sword; when applied effectively, they foster device innovation and the development of rational evaluation methods for safety and efficacy in medical devices. However, they can also form barriers for innovation. Guerra-Bretaña and Flórez-Rendón [4] emphasizes the need for a more streamlined regulatory evaluation to encourage innovation, while Onur and Söderberg [5] finds that shorter regulatory review times can lead to increased innovation in the medical device market. Stern [6] express concerns regarding the potential hindrance that regulations may pose to innovation in the medical device industry. Other authors raise concerns about the FDA approval system's ability to ensure the safety and effectiveness of complex medical devices, suggesting that regulations may slow down innovation in some cases Curfman and Redberg [7]. Further more, there are debates about implementing an even more transparent and evidenced-based system of regulation [8]. Further research and interdisciplinary collaboration are required to comprehensively evaluate the impact of regulations on medical device innovation and to develop effective policies that foster both patient safety and technological advancement.

There is still a lack of research investigating the complexity in the regulatory affairs field. Some work has been published on the linguistic complexity of EU regulations [9], but this is only part of the overall complexity experienced by stakeholders. This paper investigates the complex intricacies of regulatory affairs in the medical device industry, aiming to gain a comprehensive understanding of the topic. This research will involve conducting semi-structured interviews with regulatory affairs professionals working in the field. The aim is to obtain insights into the complexities and challenges that are faced in the regulatory field. The findings from these interviews will contribute to enhancing knowledge and identifying potential strategies for navigating any issues of regulatory affairs in the medical device industry.

## 2 Methodology

The application of qualitative research methods in system engineering has been recognized for its ability to provide nuanced insights into complex social phenomena that intersect with engineering practices [10]. Qualitative research methods, such as interviews, focus groups, and ethnographic observations, offer valuable tools to explore the social, cultural, and behavioral aspects that influence a system of systems. These methods enable researchers to uncover contextual factors, understand stakeholder perspectives, and delve into the dynamics of human interactions within a system. In the medical field, several qualitative studies have been conducted to investigate key aspects of clinical and ethical implications. For example, Dhruva et al. [11] sought expert opinions on the consequences of expedited development and regulatory review pathways for new drugs and devices. Polisena et al. [12] explored the use of real-world data and evidence for medical devices through interviews. Additionally, Zhang et al. [13] focused on identifying factors influencing patients' preferences for primary healthcare institutions, employing qualitative methods to uncover the complexities of patient decision-making. These examples show the value of integrating qualitative methodologies within other domains.

### 2.1 Research Design

This study adopted a qualitative exploratory approach, employing interviews to delve into the factors that generate perceived complexity. The investigation specifically concentrated on professionals in the field of regulatory affairs, as they are at the forefront of dealing with these complexities and often are the mediator between device manufacturers and regulatory authorities.

### 2.2 Research Ethics

Approval for this study was granted by the Ethics Committee of University of Oxford. (Central University Research Ethics Committee (CUREC) Approval Reference: R71265/RE003). The study procedures, along with the associated risks and benefits, were explained to all participants before data was collected. The interview was only conducted after informed consent was obtained.

### 2.3 Semi-structured interviews

The interview consisted of posing several key questions to identify the underlying factors that contributed to the complexity within this field. By further exploring emerging themes, a deeper understanding can be gained regarding the challenges that were encountered.

The interview study was carried out in 2023. It was conducted online using Zoom and a snow ball strategy was applied for recruitment. In total 28 volunteers participated this interview.

### 2.3.1 Inclusion/exclusion criteria

Inclusion:
1) The interviewees should hold a position directly related to regulatory affairs, such as the VP of Regulatory Affairs, CEO of a medical device company, or a Regulatory Affairs Manager or Specialist within a organisation.
2) If the participant worked for a manufacturer then the company that the participant worked for should be officially registered in accordance with the national law of the country in which the company was based.
3) The participant was involved should engage in medical device registration activities.

An overview will be provided with regards to the occupation of each volunteer, the relevant local authority they would engage with and the professional group that would best described their domain.

Table 1: Characteristics of interview participants. These include the job the volunteer was holding at the time of the interview, the regulatory authority they were engaged/involved with and the work domain that best described the participant.

| No. | Occupation | Authority | Groups |
| --- | --- | --- | --- |
| 1 | MNC Regulatory Affairs Senior Manager | FDA | Manufacturer |
| 2 | Regulatory Affairs consultancy company CEO | EU, FDA | Consultancy |
| 3 | NMPA consultant | NMPA | Regulator |
| 4 | RA manager for Boston Scientific | FDA | Manufacturer |
| 5 | VP of Regulatory Affairs | FDA, EU | Manufacturer |
| 6 | Independent consultant of regulatory affairs | FDA, EU | Consultancy |
| 7 | Regulatory Affairs Engineering | NMPA | Consultancy |
| 8 | Regulatory Affairs specialist | NMPA, FDA | Manufacturer |
| 9 | Regulatory Affairs specialist | EU | Manufacturer |
| 10 | Regulatory Affairs consultancy company CEO | FDA | Consultancy |
| 11 | MNC Regulatory Affairs Senior Manager | China | Manufacturer |
| 12 | International regulatory affairs specialist | FDA, China | Manufacturer |
| 13 | Regulatory Affairs consultancy company CEO | FDA, China, EU | Consultancy |
| 14 | VP of Regulatory Affairs | FDA, China | Manufacturer |
| 15 | Independent consultant of RA | FDA | Consultancy |
| 16 | Quality and regulatory manger | FDA | Manufacturer |
| 17 | Regulatory Company CEO | EU | Consultancy |
| 18 | Regulatory Affairs Specialist | NMPA, FDA | Manufacturer |
| | | | Continued on next page |

Table 1 continued from previous page

| No. | Occupation | Authority | Groups |
|---|---|---|---|
| 19 | Regulatory Affairs Senior Specialist | NMPA | Manufacturer |
| 20 | Regulatory Affairs Manager | FDA | Manufacturer |
| 21 | Regulatory Affairs Senior Manager | EU and FDA | Manufacturer |
| 22 | Regulatory Affairs VP | EU | Manufacturer |
| 23 | Quality and regulatory manger | FDA and EU | Manufacturer |
| 24 | Quality and regulatory manger | China | Manufacturer |
| 25 | International regulatory manger | EU, FDA | Manufacturer |
| 26 | MNC (Multinational Corporation) Regulatory Affairs Senior Manager | EU | Manufacturer |
| 27 | International regulatory specialist | FDA, EU | Manufacturer |
| 28 | Regulatory specialist | FDA | Manufacturer |

## 2.4 Data Collection

Semi-structured interviews were conducted and each interview lasted between 45 to 60 minutes. Each interview was conducted by two researchers, both knowledgeable about regulatory affairs and medical device industry, and experienced in qualitative research methods. One researcher conducted the interview, while the other diligently documented the responses and ensured the smooth flow and integrity of the interview process.

Once a participant confirmed their participation, an interview was conducted using the online platform Zoom. Only volunteers who met the inclusion criteria, expressed willingness to participate, and provided signed informed consent were ultimately chosen as interviewees.

The majority of the interviews were conducted in English. However, for a few interviews conducted in Chinese, the questions were translated on the spot from English to Chinese, and the responses were translated back into English.

The interviews were semi-structured and recorded using the Zoom software. Afterwards, the recordings were sent to a professional transcription service, where they were converted into MS Word files. The qualitative analysis data were then cleaned up and organized in the MS Word files. To facilitate our analysis, we employed Nvivo (Microsoft 12), a qualitative data analysis software, to code and categorize themes derived from the collected data. The software allowed us to systematically organize and analyze the data by assigning relevant analysis codes to specific themes.

The interview guide was developed based on the current literature. The interviewees would be asked about their perceptions and experiences related to regulatory affairs.

## 2.5 Data Analysis

### 2.5.1 Open Coding

Several recurring themes and ideas were identified during the data analysis phase. These were then summarized in a codebook using concise descriptors, referred to as nodes in NVivo. These codes were subsequently applied to code the transcripts from the interviews.

In our qualitative analysis, the coding structure was meticulously designed to facilitate a nuanced exploration of thematic elements. We initiated this process by establishing a In our qualitative analysis, the coding structure was meticulously designed to facilitate a nuanced exploration of thematic elements. We initiated this process by establishing a "parent node", representative of a broader thematic concept, such as "complexity from legal language." This node then served as a foundational point for branching out into one or more "daughter nodes", such as "lack of legal education" or "difficult interpretation." These subsidiary nodes provided a granular and detailed elucidation of the overarching parent node theme. Our methodological approach was deeply rooted in interpretive phenomenology, guiding our open coding strategy. This approach was instrumental in interpreting and understanding the lived experiences of our respondents. By employing open coding, we deliberately refrained from imposing preconceived theoretical frameworks on our data. Instead, we sought to immerse ourselves in the raw, unfiltered narratives of our participants. This approach was pivotal in uncovering and appreciating the intricate complexities embedded within the regulatory landscape. Through this meticulous process, we were able to cultivate a profound and empathetic understanding of the nuanced realities faced by those navigating this domain.

As the project progressed, we conducted a meticulous review of the codebook and the results obtained from the NVivo analysis to uncover overarching concepts. These concepts form the basis of the themes identified through thematic analysis [14], which will be presented in a subsequent section. To ensure the validity and reliability of the coding process, two independent reviewers carefully examined the codes and engaged in thorough discussions to share their perspectives. In cases where differences of opinion arose, a third reviewer was consulted to facilitate resolution and achieve consensus.

## 2.6 NLP Coding

In our study, Natural Language Processing (NLP) techniques were leveraged to delve deeper into the dataset. Specifically, we applied NLP methodologies to the transcripts derived from our interviewees, who were categorized into three distinct groups: Manufacturers, Regulators, and Consultancy. To this end, N Gram Analysis was employed on the transcripts from each of these groups, aiming to uncover and analyze underlying linguistic patterns.

Our analytical framework was augmented by the integration of machine-based methods alongside manual coding, particularly for the purpose of topic modeling. A significant aspect of this computational analysis involved the application of the Latent Dirichlet Allocation (LDA) method. Renowned for its efficacy in both natural language processing and machine learning domains, LDA serves as a probabilistic model adept at uncovering concealed thematic structures within extensive textual datasets. This approach's effectiveness in revealing hidden topics within textual data has been underscored in the literature, notably by Bhat et al.[15]. Additionally, we utilized N Gram Analysis, recognized for its utility in exposing distinctive language patterns in NLP applications, as delineated in studies such as Vatanen et al. [16].

## 3 Results

28 RA professionals volunteered and were included in the study. Notably, a number of participants showcased extensive experience across multiple regulatory domains, enhancing the study's depth. Specifically, 2 experts possessed insights into both NMPA and FDA practices, 1 expert demonstrated expertise in both NMPA and the EU, while 5 experts navigated the areas of FDA and EU collaboration. Furthermore, 1 interviewee had a background of engaging closely with NMPA, FDA, and the EU. The subsequent analysis delves into the nuanced insights shared by participants, offering a holistic view of the regulatory landscape across different jurisdictions, see Table 1.

### 3.1 Results of Opening Coding

The interviews mainly generated five key themes about contributing factors of complexity in regulatory affairs: (1) Complexity in Legal Language, (2) Complexity in Registration Process, (3) Complexity in Available Database, (4) Complexity in Product and (5) Complexity in Global Level.

Figure 1 presents a visualization of all the nodes in the project. This figure provides a concise overview of both the parent themes and the corresponding child themes that collectively capture the intricacies within regulatory affairs. We found that these child nodes were inherently connected, as one was leading to another. For example, a lack of database, which will cause trouble in mother node "registration process", also led to ambiguity in regulatory interpretation. A more detailed description of some of the most frequently mentioned nodes will be provided later in the paper.

Figure 2 showcases a word cloud generated from the responses of all the interviewees. This word cloud has been constructed based on a frequency analysis of the words used in the responses of the volunteers.

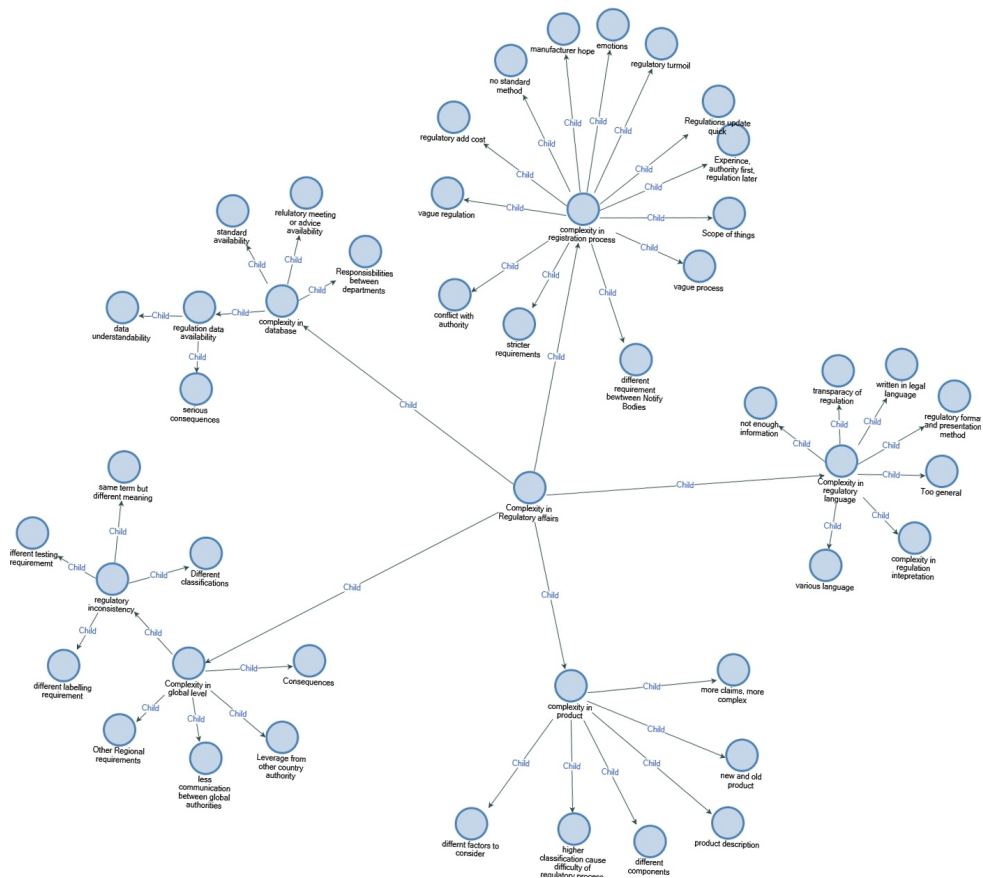

Figure 1: Complexity Nodes Using Nvivo

### 3.1.1 Interview Theme 1 : Complexity in Legal Language

During the interviews, a prevalent concern expressed by the interviewees was the perceived generality of legal language within regulatory frameworks. This generalization often resulted in challenges for manufacturers in adapting their specific devices to conform to overarching regulations. Consequently, a sense of ambiguity arose regarding the necessity of compliance with certain regulations.

"Everyone feels that no matter where the laws and regulations are, they are very obscure. We want to know whether regulations can be written more easily so we know how to implement it", said a medical device innovator. The general feelings about different country-specific regulatory environments were also different. According to interviewees, the language used in FDA regulations are relatively easier to read compared to those in the EU. As depicted in Figure 1, the child themes emanating from this mother theme comprised of the following aspects:

### 3.1.2 Obscurity of legal language

An overarching theme arising from the interviews centers on the intricate nature of legal language employed within regulatory texts. The utilization of legal terminology often gives rise to complexities and ambiguities that can be daunting for individuals without a legal background.

This issue is exacerbated by the necessity to align legal mandates with technical specifications, a task that demands a deep understanding of both domains. The use of terms such as "shall", "must" and "should" was cited as a prime example of the nuanced language that necessitates clear comprehension. Additionally, the distinction between legal norms and technical concepts further contributes to a sense of opacity, rendering interpretation an exercise that requires interdisciplinary collaboration and contextual awareness.

The legalistic style of language often leads to a communication dissonance, particularly for individuals of a more scientific background. These professionals find it challenging to discern the underlying technical requirements obscured by the legal prose, contributing to a perceived lack of clarity.

### 3.1.3 Presentation of regulations

Issues stemming from the format and presentation of regulations, hindering effective information delivery and comprehension. The presentation format of regulations can vary considerably between different regulatory bodies. In the case of the FDA, regulations are structured with a clear hierarchy, starting from the Code of Federal Regulations (CFR), followed by regulations and guidelines. Changes are made within a coherent framework, continuously refining the system. This clarity aids accessibility and understanding. Conversely, European regulations, such as MDR (Medical Device Regulation), can be extensive and complex. The MDR's comprehensive coverage necessitates section-by-section analysis, potentially hindering comprehension, according to interviewees. The interlinked nature of FDA regulations streamlines navigation and offers a more intuitive experience, while the European counterpart may require more effort to extract relevant information. Additionally, the language used in FDA regulations appears comparatively straightforward, fostering ease of comprehension, especially for non-native English speakers. This layout reduces the need for constant cross-referencing between visuals and text, as seen in the EU regulations. This clarity benefits those not fluent in English and those interacting indirectly with health authorities. In summary, the FDA's structured and coherent format, along with its language simplicity, enhances accessibility and comprehension for a broader audience compared to the more intricate European regulatory documents.

As depicted by interviewee 2: "It [the FDA] seems a bit more direct ... It's a bit easier to go through ... It seemed just easier..."

### 3.1.4 Hard to accessing sufficient information

An issue of noteworthy concern voiced by interviewees pertains to the challenge of acquiring comprehensive and timely information. It was evident that the landscape of regulatory information, especially within the context of medical devices, is subject to frequent updates and revisions. This rapid pace of change can render it difficult for stakeholders to keep abreast of the latest guidelines, leading to potential gaps in knowledge. Given the criticality of aligning devices with regulatory mandates, interviewees highlighted the necessity of dedicated efforts to stay informed through continuous engagement with regulatory agencies, industry peers, and information dissemination channels. The perceived scarcity of consolidated and readily accessible information amplifies the intricacy of compliance efforts.

### 3.1.5 Various languages of regulations

In the context of language diversity's impact on regulatory understanding, insights from the interviews revealed that European regulations are translated into various languages, yet their legalistic language often hinders comprehension for scientifically inclined individuals. The difficulties of multilingual translation were highlighted, showcasing how variations in interpretation arise due to linguistic nuances, potentially distorting regulatory intent. This complexity is akin to deciphering a foreign language, further emphasizing the importance of precise translation to maintain clarity and intended meaning across linguistic contexts. These observations collectively emphasize the relationship between regulatory language, translation, and cross-linguistic comprehension, underscoring the challenge of ensuring consistent regulatory adherence across diverse languages.

As mentioned by Interviewee 9: " … the Europeans system has multiple languages. The regulation is drafted in English, but then it's translated into the various languages. The translation to other languages creates a lot of additional trouble, because sometimes you translate sentences in a different way, and this is perceived by the readers in a different way."

### 3.1.6 Regulation interpretation

The problem of interpreting regulatory mandates emerged as a focal point within the interviews. A prevalent observation was the inherent complexity and ambiguity present within regulations, leading to divergent interpretations among professionals in the field. This multifaceted issue is compounded by the balance between regulatory precision and the need for regulatory frameworks to accommodate a broad spectrum of devices, each with its unique characteristics. As a result, manufacturers, regulators, and legal advisors are often confronted with the challenge of deciphering regulations in a manner that aligns with the intended objectives while accounting for the nuanced contexts of various devices.

Various countries have their own specific legal interpretation problem. In EU, interviewee 17 said about the MDR that "I think it is a complex regulation, and I can tell you that nobody can understand this regulation. If anybody say they can understand. I would love to see that individual. Because it's still complex. I don't think that they did due diligence when they released the MDR."

A Chinese manufacturer, interviewee 19, commented NMPA regulation: "For us, when it comes to interpreting regulations or reading regulations, the common situation is, we think we can understand all the words, but we need to further consider the intended meaning behind them. We have had very detailed communication with the regulatory drafters and some evaluators. From the perspective of the regulatory drafters, they may feel that their expressions are very straightforward. However, from our perspective, we often do not know how to proceed to the next step."

This issue was rather common across all regions. People found that the language used in medical device regulations presents various challenges, hindering smooth compliance and implementation. Regulations sometimes remain too general, lacking the necessary specificity to address the complexities of the medical device industry. This ambiguity can lead to confusion in interpreting requirements and varying understandings among stakeholders. Moreover, the complex legal language used in regulations can make it difficult for manufacturers, regulatory authorities, and healthcare professionals to precisely interpret the obligations and standards. Multiple language versions of regulations, along with diverse formats, further complicates the compliance efforts. Lack of transparency and insufficient information in some regulations can also hinder stakeholders from making well-informed decisions. Addressing these issues is crucial for promoting clarity and consistency for regulatory compliance.

## 3.2 Interview Theme 2: Complexity in Registration Process

The registration process constitutes a comprehensive continuum that spans the entirety of the medical device life cycle, encompassing phases from research and development (R&D) through market listing, and extending into the post-market phase. This procedural framework fundamentally shapes the daily operations of professionals within this domain. Its purview extends beyond scientific considerations, incorporating facets such as device classification, laboratory testing, regulatory authority assessments, clinical trials design, animal experimentation, post-market surveillance, and renewal activities [17]. Moreover, within the realm of regulatory affairs activities, the registration process provides essential administrative and commercial insights crucial for entities involved in regulatory compliance. This encompasses considerations pertinent to Marketing Authorization Holders (MAH), hiring Contract Research Organizations (CRO), as well as Good Manufacturing Practice (GMP) activities. Notably, the commercial dimension extends to strategic decisions. These should also conform to regulations regarding the in-house management of regulatory activities or potential outsourcing to external entities[18]. Below are some child nodes linked to this mother node.

### 3.2.1 Regulatory Quagmire

The interviewees feel a sense of regulatory quagmire due to higher-level regulations that are not adequately designed. As a result, some complementary measures or regulations, as well as smaller components, may not be effectively integrated into whole framework. Summarizing from these interview texts, there are certain issues with the regulatory framework of the NMPA. During the formulation of regulations, the design may not be entirely logical due to higher-level regulations, leading to difficulties in integrating complementary measures, regulations, or smaller components effectively. According to Interviewee 19, "In China, there is localized regulatory model where the provincial authority oversees products registered in that province. However, when surpassing this model, products produced overseas may only fall under the jurisdiction of the central NMPA. In this way, certain industrial-level demands currently remain unmet. Additionally, there are challenges with the production of products overseas under a domestic license, potentially creating a sense of territorial affiliation for the NMPA. Conversely, full imports enjoy considerable flexibility, and the design under the MAH system faces certain constraints. Registration requirements may seem somewhat illogical at times, reflecting issues in system settings. However, there has been [a] gradual improvement in regulations and technical documents in recent years, with various aspects gradually falling into place."

Regulatory quagmire is common across many countries. According to Interviewee 13, "Because in Europe, [a] device is not registered with an authority. You need to go to a notified body, and the notified body would need to involve an authority, and both of them need to get an approval for a device, which is very complicated." According to Interviewee 26, "So the other problem with Europe is that, in the FDA, it's a single agency for the entire United States, and in Europe there's 38 or so Notified Bodies now, it's like having 38 different possible clearance or approval organizations. So that causes a lot of disconnect."

This child theme encapsulates the prevailing sense of instability and unpredictability that industry professionals face while attempting to navigate the maze of regulatory frameworks. A central concern that emerged from the interviews is the ongoing ambiguity surrounding regulations such as the European Medical Device Regulation (MDR) and In Vitro Diagnostic Regulation (IVDR). The shifting deadlines, evolving requirements, and the dearth of clear guidance collectively create an environment of regulatory flux. This not only impedes companies in their efforts to comply but also raises concerns about potential compromises in patient safety and the timely availability of essential medical products. As a result, regulatory professionals find themselves in a perpetual struggle to adapt and realign strategies in response to this ever-changing landscape.

### 3.2.2 Rapid Regulatory Dynamics

The speed at which regulations are updated can be overwhelming for industry professionals. The continuous changes may result in confusion and uncertainty, making it difficult for manufacturers and practitioners to adapt swiftly. Industry experts shed light on the challenges posed by the swift and intricate updates to regulatory frameworks. Said the Interviewee 17, "In EU and UK, you could see some regulation changes week to week, or sometimes even within the same day. It's just hard to get a kind of a straight and a consistent answer." The European MDR and IVDR serve as primary examples, with their stipulations demanding a comprehensive understanding and swift adaptation. The interviews underscored the multifaceted nature of these regulations, which often leaves regulatory professionals grappling with the complexities that arise from this. The dynamic nature of these rules, coupled with the perpetual release of updated information, creates a pressing need for professionals to stay informed and continuously adjust their compliance strategies.

### 3.2.3 Authority Supersedes Regulation

In some instances, the expertise and authority of individuals involved may play a more dominant role in decision-making than strict adherence to regulations. This can lead to variations in interpretation and application of regulations across different entities. Interviews unveiled the pivotal role that seasoned expert and authoritative networks play in shaping regulatory compliance strategies. This transcends the influence of written regulations, as professionals often draw upon accumulated insights, historical conventions, and the guidance of respected figures in the field. The theme illuminates instances where a rigid adherence to regulatory text takes a backseat to the practical wisdom offered by those deeply entrenched in the regulatory sphere. Collaborations with experts such as medical writers, statisticians, and clinical professionals emerge as indispensable tactics to bridge the gap between regulatory mandates and practical implementation. However, concerns are raised, particularly for newcomers and startups who may face barriers in building the necessary networks and historical insights to effectively navigate this nuanced landscape.

### 3.2.4 Interpretation Discrepancies

The absence of a standardized method for interpreting regulations can lead to inconsistencies in compliance efforts, introducing varying interpretations among different stakeholders and complicating the compliance process. This lack of uniformity across diverse regulatory bodies and geographical regions emerges as a salient theme, highlighting inconsistencies in viewpoints and approaches taken by regulatory agencies. Notified bodies, responsible for assessing the conformity of medical devices in EU [19], may apply different interpretations and criteria during the certification process. This lack of harmonization can lead to varying certification outcomes, adding complexity to the regulatory landscape. The theme underscores that this lack of alignment causes confusion and frustration among companies striving for compliance, emphasizing the necessity for more cohesive communication between regulatory bodies and industry stakeholders. A collective effort to bridge these disparities is pivotal for creating a streamlined approach to regulatory adherence. Additionally, manufacturers often acknowledge that cost considerations influence their decision-making process, leading them to opt for approaches that less expensive. The regulatory process for medical devices may be characterized by ambiguity and lack of clarity, which in turn presents challenges for understanding step-by-step compliance procedures and resulting in delays and potential misinterpretations of requirements.

Interview 13, who oversees international registration department for a company: "In the context of international registrations, there can be disagreements and misunderstandings due to varying understanding of regulations. This can be influenced by factors such as cultural differences, as communication with counterparts from countries like the United States may be more trust-based, while interactions with peers from Germany may involve a meticulous examination of regulations. In the case of Chinese regulations, which often have multiple layers, conflicts and misunderstandings can arise if one party focuses on a layer that hasn't been clearly communicated. The key factors contributing to these conflicts are individual perspectives, cultural nu-ances, and the presence of biases and confidence issues. The conflicts often stem from different interpretations based on distinct levels of regulatory understanding between parties."

Another topic related to the need to understand the study design in order to provide regulators or notifying bodies with the correct evidence. Interviewee 2 mentioned; "Manufacturers sometimes present their [clinical] study by stating they want to set the trial size at N. They don't perform a sample size calculation. So, they just pick a number of patients to cut costs. In such cases, when they present their study to the regulatory authority, the authority might question them on the lack of a sample size calculation because they didn't follow a statistical approach and provide scientific evidence. Consequently, they could face challenges. For instance, if they claim their device can perform certain functions, but their clinical study doesn't support this, they will face scrutiny on the study design and it may not be accepted."

### 3.2.5 Ambiguities in the Scope of Regulatory Requirements

The extent of regulatory requirements for medical devices may not always be clearly outlined. Uncertainty regarding which specific regulations that the specific devices should fall under can hinder the proper classification and evaluation of products, posing difficulties for manufacturers seeking approval. According to Interviewee 5, "Regulations never provide you with the recipe; they always give you a framework within which you need to operate. However, they don't explicitly detail what you are required to do based on the regulation. I believe it's like a decision tree determining whether something should enter the process or not. That's where companies often face difficulties."

## 3.3 Interview Theme 3: Complexity in Available Database

### 3.3.1 Regulatory Database Availability

The interviewees also shed light on the challenges associated with regulatory database availability within the medical product sector. They highlighted that regulatory information is often dispersed across various databases, leading to inefficiencies in compliance efforts. The interviews revealed that different regions have their own regulatory frameworks and databases, which can complicate the compliance landscape. Participants expressed concerns about the lack of a centralized and comprehensive database that covers all relevant regulations. This decentralized approach makes it challenging for manufacturers to access accurate and up-to-date information efficiently. The interviews underscored the need for a harmonized and accessible regulatory database that encompasses various departments and jurisdictions, simplifying compliance processes and enabling stakeholders to navigate the regulatory requirements more effectively.

According to Interviewee 4, "[It] is sometimes more difficult to find the information. It's not that the information is not there. One of the issues is, in order to find something you have to know what you're looking for. So if you don't have your intended use correctly defined, you won't know what you're looking for. And then you have to know where to look, and how ... This is a systematic issue." And Interviewee 19; "As a consulting firm, we're still getting a lot of very basic questions that the Commission should have answered, the IMDRF should have made a bigger effort... a lot of companies like us still need to answer the paying clients questions, but they don't really help with the information sharing that's actually required industry wide. They're only very specific to our clients, and that's it. So I think there's a big deficit in how people are getting the information, and how people are sharing the information."

### 3.3.2 Standard Availability

The theme of availability of standards emerged prominently during the interviews, highlighting its vital role in the regulatory landscape of medical devices and diagnostics. Interviewees pointed out the diversity in standards across different regions and regulatory bodies, making it challenging for manufacturers to ensure compliance. This lack of uniformity poses difficulties for understanding and implementing regulatory requirements consistently. The interviews also revealed that while existing standards are valuable, there remain major gaps that require continuous development and harmonization. Participants emphasized the importance of technical experts contributing to standardization efforts, as this collaborative approach ensures that standards are comprehensive and applicable to a wide range of devices. The interviews underscored the need for greater coordination and harmonization among regulatory bodies to facilitate a cohesive and globally recognized set of standards. Interviewee 1 mentioned: "... technical experts should be contributing much more to standardization ... Standards are really valuable for lot of devices, and without a standard everyone would be lost."

## 3.4 Interview Theme 4: Complexity with Product

Numerous interviewees have noted that various products exhibit distinct levels of complexity. This theme encapsulates multiple factors influencing the regulatory compliance landscape, encompassing child nodes such as product classification, which relates higher classification with increased compliance complexity, since products with a higher classification may entail a larger set of features, components, or functions, translating into a greater volume of technical documentation and data that must be thoroughly analyzed and validated during the registration process[20]. Furthermore, it is customary for a clinical trial to be mandated in the case of higher classification devices. In contrast, for devices with lower classifications, the requisites are less stringent, with only clinical evaluation being necessary[21].

The child node on the availability and duration of relevant guidance has been released and presented to public, which play a pivotal role, as evidenced by Stern [6], highlighting that the presence of pertinent guidance significantly influences the approval timeline for medical devices. Interviewees within the industry emphasized that all stakeholders require a period to familiarize themselves with any new regulations, particularly if they represent a novel paradigm. Interviews also shed light on the temporal considerations of regulators, where the time a specific type of device has been on the market influences regulatory decisions.

The child nodes related to the claims made by a product, along with the number of components it includes, further shows how compliance can be affected by multiple factors. This is also a notion that is supported by prior research on product complexity and the interactions of components [22].

In reality there is a lot of connection between all these nodes. Interviewee 4 puts it as follows: "The more complex a design is, the harder it is to comply with the regulation, because it means that you're going to have to run more tests … if you claim that your device does something very simple and not critical, then it's much easier to follow the steps to certify it."

### 3.5 Interview Theme 5: Complexity at a Global Level

This facet of complexity is predominantly pertinent to global manufacturers aiming to expand market reach across different regions. The expensive nature of medical technology development also often requires manufacturers to consider multiple markets. As shown in Figure 1, particular emphasis is placed on the following child nodes below.

The node of "Regulatory Inconsistency", describes divergent labeling or testing requirements, among other variations. This leads to discrepancies in regulatory decisions and information supplied to medical practitioners and patients[23], which can pose risks to safety and effectiveness[24]. Regulatory inconsistency were mentioned by multiple interviewees and it has been a major concern if they wanted to move between markets. Interviewee 3 stated: "Unfortunately, the further we get into MDR for the EU, the less harmonization there is even amongst the platforms in the EU …" This highlights the issues that even occur within a specific region, let alone the differences that exists between e.g. the US, Japan, China, EU and other countries. Similar occurrences are noted in other geographical contexts as well. A Regulatory Affairs Specialist, employed in the Brazil branch of a multinational company serving both Brazil and the USA, expressed the following when talking about the difference that exists between a notified body, health authority or a given ISO standard: "So who should I follow? If they're different, … for example, in Brazil, if I have a ISO 13485 certification, it does not exempt me from needing a GMP from ENVISA, since they're not the same. The ISO certificate is not a requirement for ENVISA, so should I follow what ENVISA dictates, or should I go with ISO? It's quite confusing for us." According to this professional's comments, it's evident that there are various certifications required across different regions. Some of these certifications might appear to serve similar functions, yet Regulatory Affairs specialists are compelled to obtain them separately. It becomes a source of frustration when certifications with ostensibly similar functions entail different details.

### 3.6 Code analysis with NLP

#### 3.6.1 N Gram Analysis

We conducted N-gram analysis on transcripts to discern potential indicators related to the three stakeholder perspectives. Our examination involved categorizing inter-viewees into three distinct groups and then completing the analysis for each group individually: Regulators, Manufacturers and Consultants, as shown in Table 2. Certain words emerge with high frequency across the groups, such as "regulatory affairs", "medical devices", "clinical trials" and "guidance documents". For the top N-grams for regulators, the focus was on terms such as "regulatory affairs" , "clinical trials", and "guidance documents". These terms underscore a regulatory focus, with an emphasis on compliance. Additionally, terms like "research development", "technical requirements", and "technology transfer" suggest a strong orientation towards research and technological aspects within the regulatory framework. The N-gram analysis for manufacturers unveils a prominent emphasis on terms like "regulatory affairs", "medical device" and "clinical trial." This emphasizes a potential dual commitment to regulatory compliance and product development. Noteworthy is the recurrence of terms like "make sure", "notified bodies" and "quality assurance," indicating a stronger focus on ensuring product quality and compliance with regulatory standards. As for consultants, frequent terms like "medical device", "regulatory affairs" and "United States" . Notably, terms such as "United States", "class III" and "registration process" suggest a specific focus on regulatory classifications and registration procedures.

Table 2: Top 10 N-grams and their frequencies for Regulators, Manufacturers, and Consultants

| Regulators | | Manufacturers | | Consultants | |
| --- | --- | --- | --- | --- | --- |
| **N-gram** | **Freq.** | **N-gram** | **Freq.** | **N-gram** | **Freq.** |
| regulatory affairs | 10 | regulatory affairs | 47 | medical device | 51 |
| clinical trials | 10 | medical device | 44 | regulatory affairs | 37 |
| guidance documents | 6 | clinical trial | 31 | united states | 35 |
| research development | 6 | guidance documents | 24 | clinical trials | 32 |
| technical requirements | 6 | make sure | 20 | conduct clinical | 30 |
| conduct clinical | 6 | medical devices | 19 | notified body | 29 |
| language used | 6 | things like | 19 | medical device | 27 |
| technology transfer | 5 | notified bodies | 17 | class III | 21 |
| regulatory authorities | 5 | feel like | 17 | registration process | 20 |
| medical products | 4 | quality assurance | 17 | notified bodies | 20 |

### 3.6.2 Latent Dirichlet Allocation Model

The Latent Dirichlet Allocation (LDA) model [25] is an approach for clustering items into distinct groups or topics. We employed this model on our transcript data and discovered that several terms, such as "product," "regulatory," "device," "FDA," and "regulation," frequently emerge in these thematic clusters.

In our study, we meticulously determined the most appropriate number of topics for analysis. This was achieved through the construction and evaluation of multiple Latent Dirichlet Allocation (LDA) models, each varying in the number of topics (denoted as k). The efficacy of these models was assessed based on their coherence scores, following the methodology outlined by Hasan et al. (2021) [26]. The model exhibiting the optimal balance, characterized by the highest coherence score, encompassed seven distinct topics. This finding is graphically represented in the model output (Figure 5).

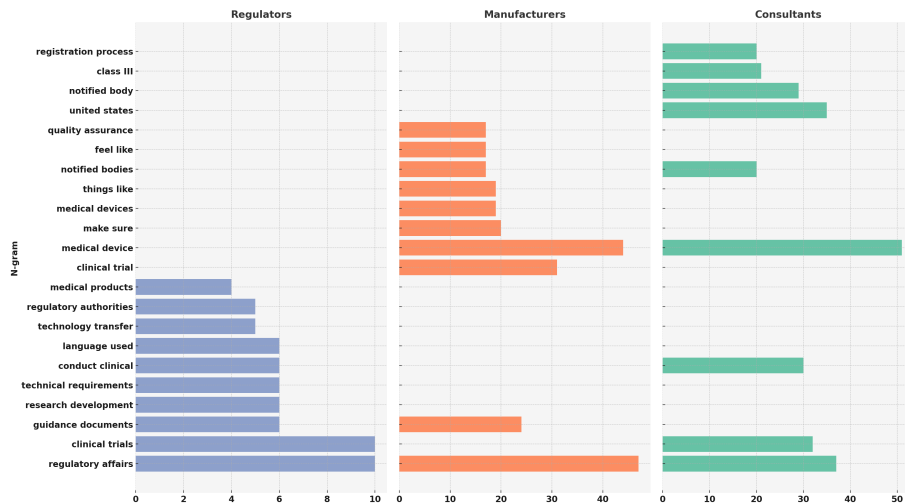

Figure 3: Top 10 N-grams and their frequencies of Regulators, Manufacturers and Consultants

Leveraging the Gensim Python library (Řehůřek & Sojka, 2011) [27], we generated an Inter-Topic Distance Map, as depicted in Figure 4. This visualization elucidates the thematic structure uncovered by our LDA models, illustrating seven discrete circles that represent clusters of words corresponding to each identified topic. A closer examination of this map highlights the relational proximity of Topic One to Topics Three and Four. Within Topic One, we observed a predominant frequency of specific terms, including "product," "device," "FDA," "regulatory," amongst others. The comprehensive list of frequently occurring terms across all seven topics is systematically presented in Table 3.

In our analytical process, we identified and extracted the most frequently occurring terms from each topic cluster. These terms were then meticulously organized to represent the thematic essence of each respective topic. Utilizing the lexical characteristics of these predominant terms, we assigned descriptive titles to each topic cluster, encapsulating their thematic focus. The following delineation presents this thematic structuring: on the left, we list the designated topic names, reflective of the central theme of each cluster; on the right, we enumerate the most frequently occurring terms within each topic, serving as the basis for their thematic classification as shown in Table 3.

## 4 Discussion

Regulations can affect various aspects of product approval and compliance. They are an essential component of the medical device industry. Due to various reasons, ex-perts have found that regulations are becoming more complex, as shown by previous research [9]. More importantly, poor regulation does increase the cost for industry and weakens the innovation process[28]. Understanding complexity is therefore essential, as it could aid decision making during the product approval process. Complexity in a regulatory environments has been previously investigated in other domains, such as financial regulation Easley and O'Hara [29], where they looked at the ambiguity of regulations. Muchmore [30] developed an analytic framework for understanding the role of uncertainty in regulatory design. More recent ideas suggest the existence of multiple dimensions of complexity within the medical device regula-tory landscape. These complexities, if left unaddressed, could hinder the development, approval, and market access of medical devices [31]. Understanding and mitigating these complexities are essential for creating a conducive environment for innovation and ensuring timely and safe access to medical technologies for patients.

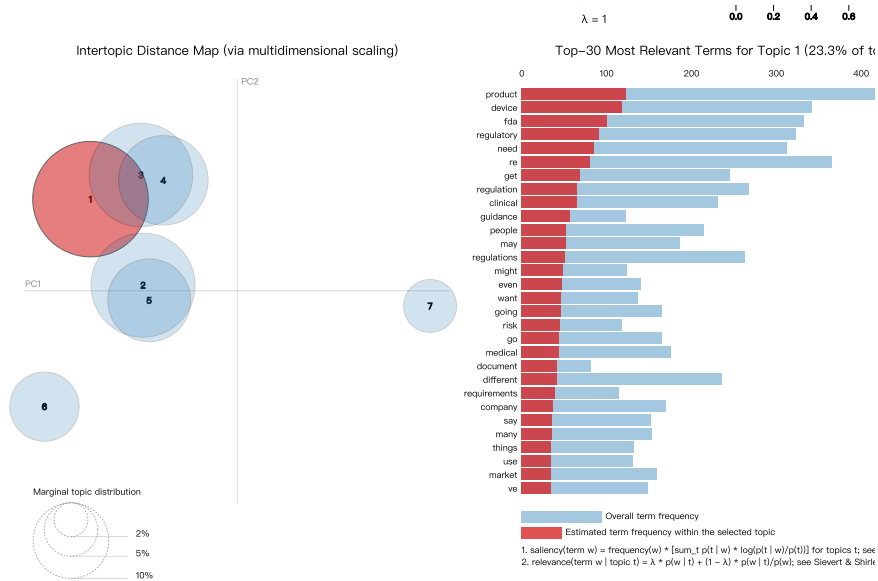

Figure 4: Inter-topic Distance Map

| **Topic** | **Most frequency Norms** |
|---|---|
| Topic 1: Product Compliance and Regulatory Oversight | product, device, FDA, regulatory, need |
| Topic 2: Diverse Product Regulation and Compliance | product, different, FDA, right, regulations |
| Topic 3: Clinical Data Analysis and Device Requirements | people, get, data, FDA, regulatory |
| Topic 4: Regulatory Landscape for Medical Devices | product, get, people, regulations, device |
| Topic 5: Broad Regulatory Framework in Clinical Context | product, regulation, devices, clinical, need |
| Topic 6: Regulatory and Clinical Product Dynamics | regulatory, product, need, device, clinical |
| Topic 7: Standards and Compliance in Device Manufacturing | device, FDA, need, standard, product |

Table 3: Results of Topic Modelling and Most Frequency Norms

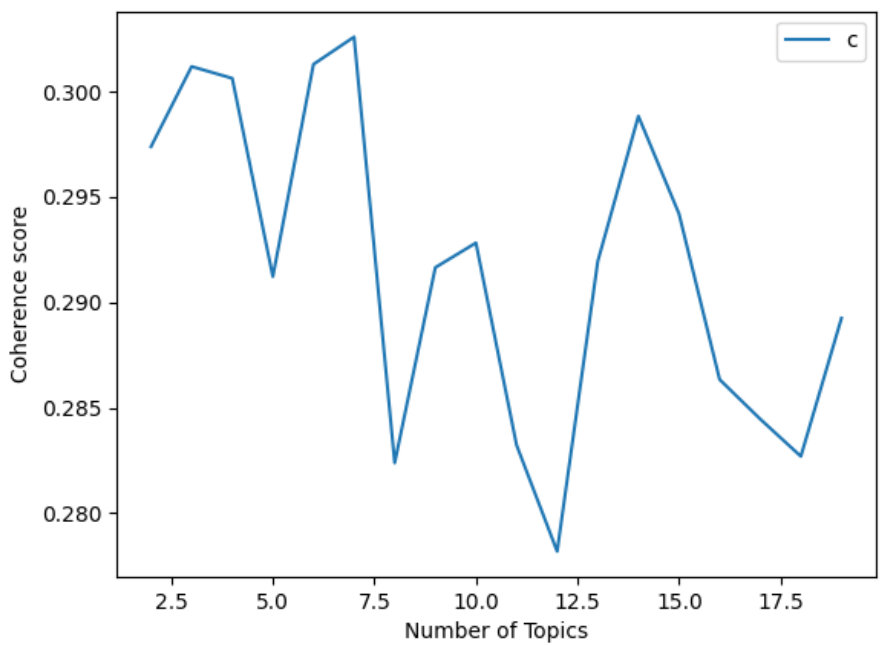

Figure 5: Find Optimal Number of Topics in Latent Dirichlet Allocation (LDA)

Our investigation revealed five primary origins of complexities that warrant consideration for mitigation: legal language complexity, registration process intricacies, database complexities, product-level intricacies, and complexities on a global scale. The coding structure involved a hierarchy, with child nodes extending from parent nodes. For instance, a lack of an appropriate database can catalyze complexity in the registration process and impedes progress due to insufficient information. This underscores that the challenges in medical device regulatory affairs constitute a system of systems problem, rather than a singular issue [32]. The escalating complexity of modern systems further exacerbates these challenges, necessitating additional research and the development of innovative approaches to effectively address these intricate issues.

The N-gram analysis unveiled nuanced distinctions in language usage across various stakeholders. Moreover, this analytical method could offer actionable insights that can help facilitate communication between different stakeholder and foster collaborative solutions tailored to the unique challenges and opportunities within each of the areas of interest.

In the realm of NLP and open coding, particularly when examining LDA-generated themes (topic names), we find notable parallels with themes unearthed by the manual coding approach. Take, for example, Topic 2, "Diverse Product Regulation and Compliance," which mirrors the theme of product complexity identified with the manual coding. In a similar vein, Topic 3, "Clinical Data Analysis and Device Requirements," echoes the theme of registration process complexity. Moving to Topic 6, "Regulatory and Clinical Product Dynamics", this resonates with the rapidly evolving dynamics we identified in open coding themes. Yet, when relying solely on LDA outcomes, it's challenging to trace these results back to open coding themes, as the latter tend to be more nuanced and explanatory. Whilst NLP findings provide valuable insights and guideposts for additional manual coding, it's crucial to recognize that, with the current state of technology, human-led manual coding remains more accurate and indispensable [33].

The European Union's medical devices framework strives to establish a harmonized and robust system that encourages innovation while prioritizing public health. This framework outlines a comprehensive set of requirements and guidelines for manufacturers to adhere to, ensuring the quality and performance of medical devices [34]. Within the EU, the regulatory landscape for medical devices is governed by specific directives and regulations. The engagement of notified bodies, independent third-party organizations designated by regulatory authorities, is crucial in the conformity assessment process [35]. However, insights garnered during the transcript process reveals notable concerns and issues. The primary problem raised is the inconsistency among the procedures and requirements for each notified body [36]. Interviewees also express apprehension about the extended waiting times for scheduled testing with testing labs in the EU, indicating that, currently, it takes years to secure a testing slot.

Despite the pivotal role regulations play in shaping medical device development and distribution, there exists a gap in research exploring the specific challenges and opportunities they pose. Further investigation is imperative to gain insights into the intricacies and implications of regulatory requirements on the medical device industry. An intriguing avenue involves integrating regulatory affairs with data science [37], offering a promising approach to streamline regulatory processes. Through the application of data tracking and analysis techniques, compliance procedures can be optimized, enhancing efficiency and effectiveness. This collaborative synergy between regulatory affairs and data science holds the potential for smarter regulation of intelligent medical devices, fostering innovation while upholding safety standards. As the landscape evolves, with more AI products emerging, regulatory affairs are also set to transform into AIRA (Artificial Intelligence Regulatory Affairs). [38] delineates several directions in which regulatory affairs are anticipated to evolve, encompassing real-time data submission to regulatory agencies via cloud-based systems, the application of digital twins in clinical trials, and Authority Real-Time review programs.

With the advent of Large Language Models (LLMs)[39], there's a burgeoning potential in regulatory affairs to utilize these AI tools for medical products. Initially demonstrated in studies using Natural Language Processing (NLP) to gauge interviewee opinions, the future could see LLMs capturing broader stakeholder opinions on regulatory changes in the medical product sector. This approach promises to reveal diverse group perspectives on regulatory modifications, fostering regulatory processes that resonate more effectively with stakeholder needs and advancing informed, responsive regulatory frameworks. In light of these advancements, it becomes crucial to discern the current complexity roadblocks in the regulatory affairs landscape, offering an opportunity to apply various AI techniques to address these challenges.

## 5  Limitation

The inherent nature of qualitative research often involves working with smaller sample sizes [40]. The results derived from our study are intrinsically linked to the specific contexts and experiences of our participants. While our participant cohort encompasses a geographically diverse array from the United States, European Union, and various Asian countries, providing a broad cultural perspective. The insights and themes extrapolated from our interviews reflect the viewpoints and experiences of this particular group, and caution should be exercised when attempting to extrapolate these findings to a wider population. This limitation underscores the necessity for additional research, potentially incorporating a more extensive and varied participant base, to further validate and expand upon our findings.

Additionally, our exploration is confined to employing two natural language processing (NLP) methods, which are the N-gram and LDA modeling. Divergent results may emerge when employing alternative NLP methodologies, underscoring the importance of recognizing potential variations in analytical outcomes among different researchers. While BERT stands out for its efficacy on larger datasets[41], our study, constrained by the available data, exclusively utilizes N-gram and LDA modeling to better fit the size of our dataset. A prospective avenue for research involves the consideration of BERT, particularly with an expanded dataset in the future.

To our knowledge, this is the first qualitative study that explores factors to regulatory affairs complexity. This study has some limitations that can be addressed in future studies. Foremost, the research primarily focus on US, EU and China, potentially limiting the generalizability of findings across a broader spectrum of regulatory contexts. Thus, future studies can be extended to other areas. Subsequently, the qualitative nature of this study allows for a comprehensive exploration of regulatory complexities, yet quantifying the prevalence and impact of specific challenges may require a complementary quantitative approach. Utilizing quantitative surveys or metrics could offer a more precise understanding of the relative significance of identified complexities. Therefore, future researches can use quantitative design to explore the influencing degree of these factors on complexity of regulatory affairs. Lastly, the study focuses on the present state of regulatory affairs and the challenges they entail. However, regulatory environments are subject to evolution and reform, potentially influencing the nature and intensity of complexities over time. To capture the dynamic nature of regulatory affairs, longitudinal studies could offer insights into trends and changes within this intricate landscape.

## 6 Data Availability Section

The data that support the findings in this study is available on: https://github.com/Oxford-NIL/Uncovering-Regulatory-Affairs-Complexity

## 7 Conflict of Interests

JB is co-founder of RegMetrics which offers a software based solution to regulatory navigation. The other authors report no conflict of interest in this research.